\documentclass[usenatbib,usedcolumn]{mn2e}
\usepackage{psfig}
\usepackage{amsmath}

\def\lapp{\ifmmode\stackrel{<}{_{\sim}}\else$\stackrel{<}{_{\sim}}$\fi}
\def\gapp{\ifmmode\stackrel{>}{_{\sim}}\else$\stackrel{>}{_{\sim}}$\fi}

\title[Interstellar scintillation in PSR~B0329+54]
     {Daily Observations of Interstellar Scintillation in PSR~B0329+54}
\author[N. Wang et al.]{ N.~Wang,$^{1}$\thanks{Email:na.wang@uao.ac.cn}
 Z. Yan,$^{1,3}$ R.~N.~Manchester,$^{2}$
 and H. X. Wang,$^{1}$\\
$^{1}$Urumqi Observatory, NAOC, 40-5 South Beijing Road, Urumqi, Xinjiang, China, 830011\\
$^{2}$Australia Telescope National Facility, CSIRO, PO Box 76, Epping, NSW 1710, Australia\\
$^{3}$Graduate University of Chinese Academy of Sciences, 19A Yuquan road, Beijing, China, 100049
}
\begin{document}

\date{\today}

\maketitle

\label{firstpage}

\begin{abstract}
Quasi-continuous observations of PSR B0329+54 over 20 days using the
Nanshan 25-m telescope at 1540 MHz have been used to study the effects
of refractive scintillation on the pulsar flux density and diffractive
scintillation properties. Dynamic spectra were obtained from datasets
of 90 min duration and diffractive parameters derived from a
two-dimensional auto-correlation analysis. Secondary spectra were also
computed but these showed no significant evidence for arc
structure. Cross correlations between variations in the derived
parameters were much lower than predicted by thin screen models and in
one case was of opposite sign to the prediction. Observed modulation
indices were larger than predicted by thin screen models with a
Kolmogorov fluctuation spectrum. Structure functions were computed for
the flux density, diffractive timescale and decorrelation
bandwidth. These indicated a refractive timescale of $8\pm 2$~h, much
shorter than predicted by the thin screen model. The measured
structure-function slope of $0.4\pm 0.2$ is also inconsistent with
scattering by a single thin screen for which a slope of 2.0 is
expected. All observations are consistent with scattering by an
extended medium having a Kolmogorov fluctuation spectrum which is
concentrated towards the pulsar. This interpretation is also consistent
with recent observations of multiple diffuse scintillation arcs for
this pulsar.
\end{abstract}


\section{Introduction}
In the early days of pulsar observations, slow fluctuations of pulsar
amplitudes were reported by \citet{chp70} and \citet{hth73}. But it
was not recognized that they were an interstellar propagation effect
until a close correlation between the time-scale of the intensity
fluctuations and the pulsar dispersion measure was pointed out by
\citet{sie82}. Subsequently, \citet{rcb84} explained these slow
fluctuations as refractive interstellar scintillations (RISS). This
phenomenon is produced by the same spectrum of electron density
fluctuations in the interstellar medium that is responsible for the
well known diffractive scintillation (DISS). DISS is caused by the
small spatial scale density fluctuations ($10^{6}-10^{8}$~m) and it
appears as intensity variations in both the time and frequency domains
with characteristic scales of minutes to hours and kHz to MHz
respectively. On the other hand, large-scale inhomogeneities
($10^{10}-10^{12}$~m) in the interstellar medium give rise to
focussing effects observed as RISS. A good review of the theory behind
this interpretation is given by \citet{ric90}.

It is generally accepted that the spectrum of electron density
fluctuations in the interstellar medium has a power-law form:
\begin{equation}
    \centering
    \label{eq:spectrum}
    P_{\rm 3n}(q) = C_{\rm n}^2 q^{-\beta},
\end{equation}
where $q=2\pi/L$ is the wave-vector associated with a spatial size
$L$. The quantity $C_{\rm n}^2$ is a measure of turbulence along a
particular line of sight. The power-law index $\beta$ which indicates
the steepness of the inhomogeneity spectrum is in the range of
$3<\beta<5$. But the details of the spectrum, e.g. its slope and the
range of scale sizes over which it is valid, still remain open
questions. Also, different assumptions are made about the distribution
of scattering material along the path to the pulsar. Many observations
are well accounted for by models in which the scattering is
concentrated in a single centrally located thin screen, but
others imply that there are many scattering sites or a statistically
uniform screen extending over most or all of the path. 

A range of observations suggest that a Kolmogorov fluctuation spectrum
with $\beta=11/3$ is widely applicable in the interstellar medium
\citep{ars95}. However, observational studies showing, for example,
persistent drifting bands in dynamic spectra and increased modulation
of diffractive properties suggest that excess power is often seen at
low spatial frequencies, implying a spectral index $\beta \gapp
4$. Examples of extensive studies are those by \citet{grl94}, who
studied eight pulsars at 408 MHz using the Lovell telescope and
interpreted changing diffractive patterns in dynamic spectra in terms
of refractive effects, and those by \citet{brg99a,brg99b} and
\citet{bgr99}. These latter authors used the Ooty radio telescope at
327 MHz to study 18 pulsars over a 2.5-year dataspan and showed that,
while many observations were consistent with a Kolmogorov spectrum,
others showed evidence of a steeper spectrum. \citet{ssh+00} reported
on 5 years of daily monitoring of the flux density of 21 pulsars at
610 MHz, showing that most of the results, covering a wide range of
scattering strengths, were consistent with a Kolmogorov fluctuation
spectrum. In some pulsars, though modulation indices were higher than
expected, possibly due to the presence of an ``inner scale'', a cutoff
in the fluctuation spectrum at scales greater than the diffractive scale
$s_d \approx \lambda/\theta_{\rm d}$, where $\lambda$ is the
wavelength and $\theta_{\rm d}$ is the angular size of the scattering
disk.

Other observations suggesting steeper power-law spectra ($\beta\gapp
4$) include fringing in dynamic spectra implying multiple imaging
\citep[e.g.,][]{cpl86,cw86,rlg97} and extreme scattering events
\citep{lrc98}. \citet{bn85} suggested a steeper spectrum to deal with
the theoretical difficulties in supporting a turbulent cascade in the
Kolmogorov spectrum. Based on power-law models of the spectrum,
\citet{rnb86} analysed the effects of RISS on diffractive parameters
for different slopes $\beta=11/3$, 4 and 4.3, predicting
anti-correlations for ($F$,$\Delta\nu_{\rm d}$) and ($F$, $\Delta
t_{\rm d}$), where $F$ is flux density and $\Delta\nu_{\rm d}$,
$\Delta t_{\rm d}$ are the diffractive scintillation bandwidth and
timescale, respectively and a postive correlation for
($\Delta\nu_{\rm d}$,$\Delta t_{\rm d}$), with the magnitudes of the
correlation coefficients being larger for steeper spectra.

More recent observations have tended to concentrate on detailed
studies of individual pulsars \citep[e.g.,][]{sss+03,rdb+06,sss+06} or
on the fascinating ``scintillation arcs'' which are seen in secondary
spectra, that is, two-dimensional Fourier transforms of dynamic
spectra, in high-sensitivity, high-resolution observations
\citep{smc+01,hsb+03,sti06}. These arcs, which are closely related to
multiple imaging and the frequently observed ``criss-cross'' sloping
bands, result from interference between rays in a central core and
rays from an extended scattering disk. They form a powerful probe of
structure in the interstellar medium \citep{crsc06}. For
example, \citet{ps06a} present recent observations of six pulsars
showing multiple arcs of different curvature implying a distribution
of scattering centres along the line of sight.

PSR B0329+54 is one of the strongest pulsars known and its
scintillation properties have been studied by many
authors. Observations at 610 MHz by \citet{sfm96} showed that the
correlations between variations of flux, decorrelation bandwidth and
scintillation time-scale were consistent with the theoretical
predictions with $\beta<4$. However, \citet{brg99b} found that for PSR
B0329+54, correlations between these parameters at 327 MHz were not in
accord with the predictions. Long-term flux density monitoring at 610
MHz by \citet{ssh+00} showed that the modulation index was somewhat
higher than expected for a Kolmogorov spectrum, i.e. $\beta >
11/3$. \citet{sss+03} took data obtained over a wide range of
frequencies finding a value of $\beta =3.5$, less than but consistent
with a Kolmogorov turbulence spectrum, and interpreted this result in
terms of weak plasma turbulence in the interstellar
medium. Observations at 1540 MHz using the Nanshan telescope
\citep{wmj+05} showed inconsistencies in values of the power-law index
obtained from two different approaches.  The ratio of refractive
scattering angle $\theta_r$ and diffractive angle $\theta_d$ gave
$\beta \approx 3.6$, while the frequency dependence of scintillation
parameters gave $\beta>4$. These conflicting results suggested
undertaking observations with a better sampling of the refractive
variations. These have a predicted timescale
\begin{equation}
\Delta t_{\rm r} \approx (2\nu / \Delta\nu_{\rm d})\Delta t_{\rm d},
\label{eq:tref}
\end{equation}
\citep{grc93,ssh+00}, about four days for this pulsar.

In this paper, we present the results of quasi-continuous 1540~MHz
observations of PSR B0329+54 over a 20-day interval in 2004,
March. Diffractive parameters were sampled at 90-min intervals, giving
good resolution of the expected RISS variations. The layout for the
rest of this paper is as follows: in Section 2 we introduce the
observations; data analysis and results are shown in Section 3, and in
Section 4, we discuss the results and their
interpretation. Conclusions are presented in Section 5.

\section{Observations}
The observations were made using the Nanshan 25~m radio telescope with
a central frequency of 1540 MHz.  Our system is sensitive to
orthogonal linear polarizations and the system temperature for the two
channels is 20~K and 22~K respectively. Each polarization has 128
channels, each of bandwidth 2.5~MHz, giving a total bandwidth of
320~MHz. After detection and high-pass filtering, the data are 1-bit
sampled and folded at the apparent pulsar period. Signals from the two
polarizations are summed to form total intensity pulse profiles.

We made our observations from March 12 to March 31 in 2004, and
obtained more than 150 scintillation dynamic spectra for PSR
~B0329+54.  The observations were continuous (PSR B0329+54 is
circumpolar at Nanshan) apart from an interruption of about 3 days
from March 20 due to telescope problems. In our analysis, the time
block for each dynamic spectrum is typically 90 minutes with a
sub-integration time of 60~s. We calibrated the pulsar flux density
scale using observations of ten strong, relatively distant pulsars
which all have well-measured flux densities in the ATNF pulsar
catalogue (www.atnf.csiro.au/research/pulsar/psrcat).

\section{Data Analysis and Results}
\subsection{Scintillation Dynamic Spectra}
\begin{figure*}
    \centerline{\psfig{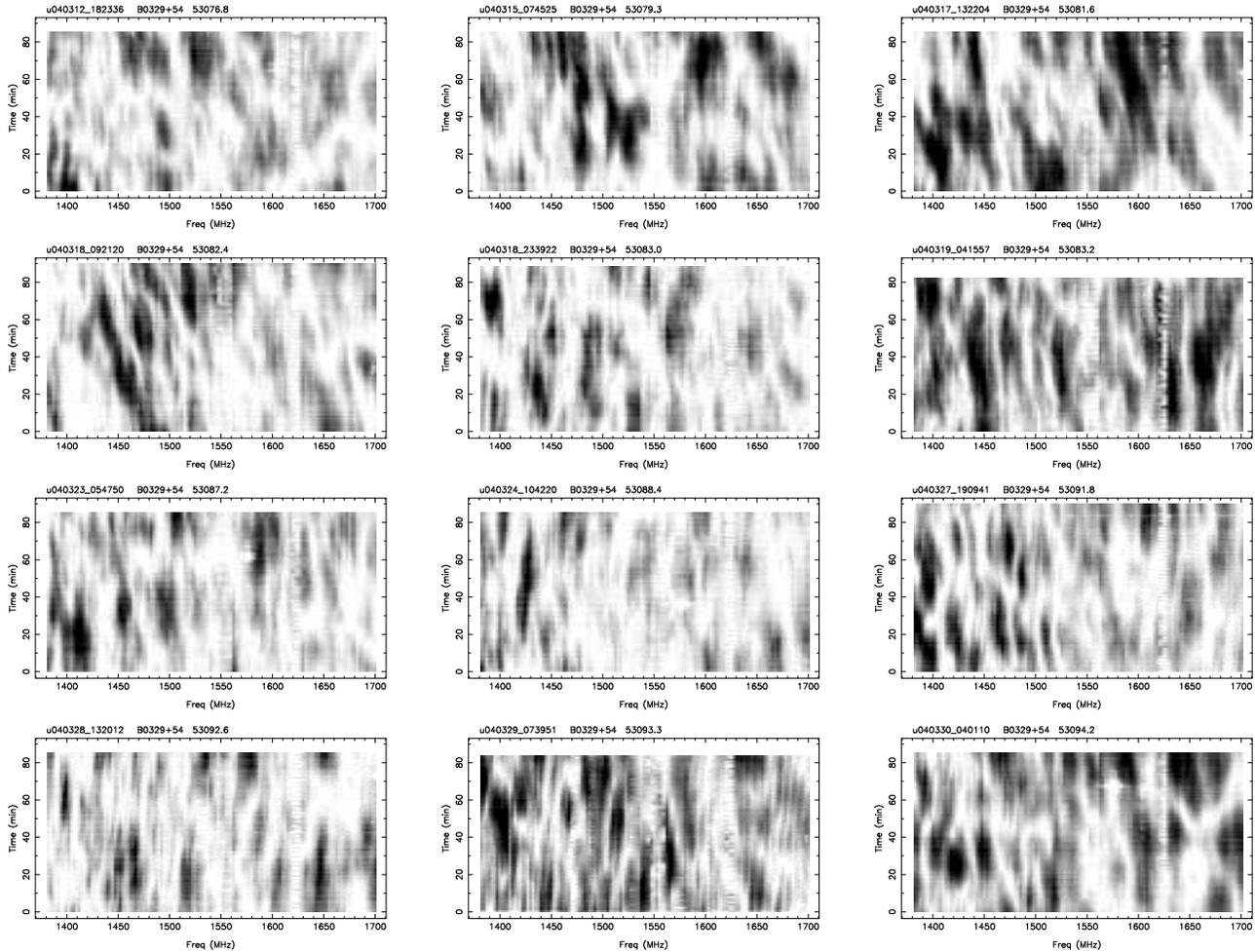}}
    \caption{Samples of dynamic spectra for PSR B0329+54 taken from
      March 12 to March 31 in 2004. The horizontal and vertical axis
      corresponds to frequency and time respectively. Darker regions
      of the grey-scale saturate at black for intensities above 85
      percent of the maximum, white corresponds to below 1 percent of
      the maximum intensity and the grey-scale is linear between these
      limits.The Modified Julian Date (MJD) of the observation is
      shown in the top right corner of each panel. }\label{fg:dynamic}
\end{figure*}

Dynamic spectra were obtained for each observation by plotting the
mean flux density in each channel against time. Examples of these
plots are shown in Fig.~\ref{fg:dynamic}. Channels badly affected by
radio-frequency interference have been interpolated over from adjacent
frequency channels.

Fig.~\ref{fg:dynamic} shows that the dynamic spectra are well
resolved and that the scintles or peaks of flux density change their
shapes and sizes in the frequency and time domains. At some epochs,
the pulsar seems weaker and the scintles are smaller (MJD 53088.4),
and at other epochs, the pulsar signal is stronger and the scintles
larger (e.g. MJDs 53081.6, 53083.2, 53093.3). The frequency dependence
of DISS can be seen within our receiver bandwidth, i.e., scintles are
smaller at the lower end of the observed band (MJD 53091.8). At some
epochs the spectra show a significant drifting pattern (e.g. MJD
53082.4), which implies a modulation by RISS. However, we didn't
observe the fine fringes seen in the data of \citet{wmj+05}. The
scintles are generally smaller in this observing session, resulting in
a smaller value in $\Delta \nu_{\rm d}$, which we will discuss more in
Section~\ref{sec:gauss}. Deep modulations are common in all the
dynamic spectra we have recorded.

The limited number of scintles in the dynamic spectra introduces a
statistical estimation error in the scintillation parameters
\citep{cwb85}. An approximate estimate of the number of scintles $N$
in the dynamic spectra is given by:
\begin{equation}
    \centering
    \label{eq:scintno}
    N = \frac{{T_{\rm obs}  \times BW_{\rm obs} }}{{\Delta \nu _{\rm d}  \times \Delta t _{\rm d} }},
\end{equation}
where $T_{\rm obs}$ is the total observing time and $BW _{\rm obs}$
the total observing bandwidth, $\Delta \nu_{\rm d}$ decorrelation
bandwidth and $\Delta t_{\rm d}$ is the DISS timescale. The fractional
estimation error is then given by:
\begin{equation}
    \centering
    \label{eq:staerr}
    \sigma_{\rm est} = \Big (0.5\times \frac{T_{\rm obs}\times BW_{\rm obs} }{\Delta\nu _{\rm d} \times \Delta t _{\rm d} }\Big )^{-0.5}
\end{equation}
where we assume a scintle filling factor of 0.5 \citep{brg99a}. This error
contribution is taken into account in the estimates of scintillation
parameters in Section~\ref{sec:gauss}.

\subsection{Secondary Spectrum}
The two-dimensional Fourier spectrum of the primary dynamic spectrum
is often referred to as the secondary spectrum. To show the beautiful
arc structures \citep[e.g.,][]{sti06}, high frequency resolution and
high sensitivity are required. Unfortunately, the Nanshan system does
not provide these. However, earlier observations
at 1540~MHz for this pulsar by \citet{wmj+05} showed sloping fringes in
the dynamic spectra resulting in offset features in the secondary
spectra. Surprisingly, despite using the same system at Nanshan,
our three-week consecutive observations show no significant arc
structure or offset features in the secondary spectra.

\subsection{Two-Dimensional Auto-Correlation Function}
To quantify the diffractive parameters, we use the two-dimensional
auto-correlation function (ACF), which was computed for frequency
lags up to half of the observing bandwidth and for time lags up to
half of the observing time. The ACF of the dynamic spectra $A(\Delta \nu,\Delta t)$
is defined as:
\begin{equation}
    \centering
    A(\Delta \nu ,\Delta t ) = \sum\limits_{\rm \nu} ^{} {\sum\limits_{\rm t}^{}
    {\Delta F(\nu ,t)\Delta F(\nu  + \Delta \nu ,t + \Delta t )} },
\end{equation}
where $\Delta F(\nu, t) = F(\nu ,t) - \overline F$, and $\overline F$ is the
mean pulsar flux density over each observation, then the normalized ACF is:
\begin{equation}
    \centering
    \rho (\Delta \nu ,\Delta t ) = A(\Delta \nu ,\Delta t )/A(0,0).
\end{equation}
\begin{figure*}
    \centerline{\psfig{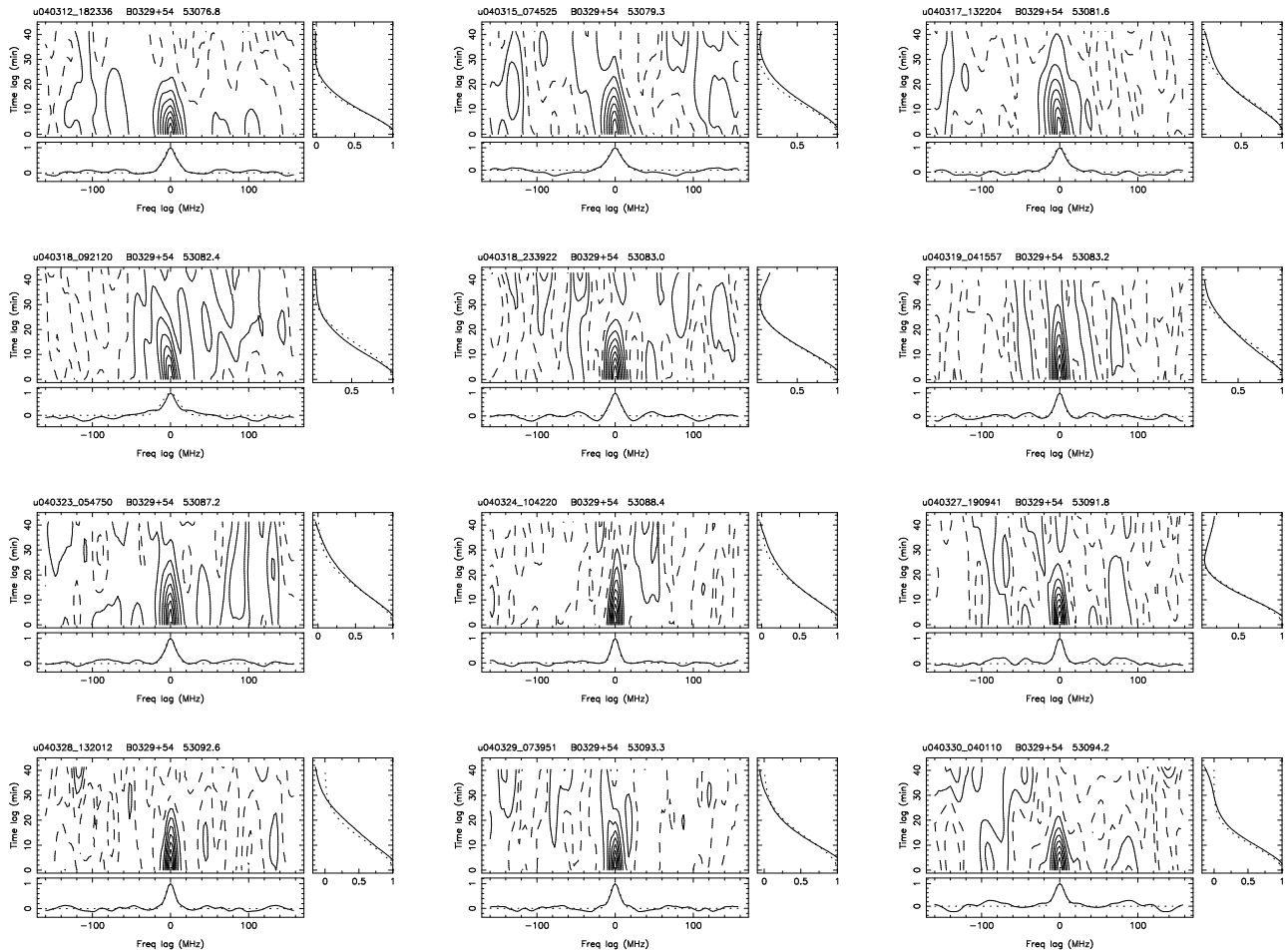}}
    \caption{Contour  plots  of  ACF  for  dynamic  spectra  shown  in
    Fig.~\ref{fg:dynamic}. There are 10  contours over the range zero
    to unity, successive contours are separated by an interval 0.1 and
    the dashed contours represent  negative values. The observing MJD
    is shown in the top right corner of each panel.  }
    \label{fg:contour}
\end{figure*}

We plot the normalized ACF and one dimensional cut at zero time and
frequency lag in Fig.~\ref{fg:contour}.  These figures correspond to
the dynamic spectra shown in Fig.~\ref{fg:dynamic}. Following
convention, the DISS time-scale $\Delta t_{\rm d}$ is defined as the time
lag at zero frequency lag where the ACF is 1/e of the maximum, and the
decorrelation bandwidth $\Delta \nu_{\rm d}$ is defined as the
half-width at half-maximum of the ACF along the frequency lag axis at
zero time lag \citep{cor86}. For weaker pulsars, random system noise
results in a strong spike at the zero time and frequency lag of
ACF. For PSR B0329+54 the spike is weak but still observable at some
epochs. To remove the spike at the origin we used the four neighbouring
frequency-lag points to fit for a parabola across the zero frequency
lag. The central point is interpolated from this fit.

\subsection{ACF and Scintillation Parameters}\label{sec:gauss}
Following \citet{grl94} and \citet{brg99a}, we use a two-dimensional
elliptical Gaussian function to fit the ACF with form:
\begin{equation}
    \centering
    \rho (\Delta \nu, \Delta t ) = C_0 \exp [ - (C_1 \Delta \nu ^2  + C_2 \Delta\nu \Delta t
    + C_3 \Delta t^2 )],
\end{equation}
in which $C_0$ is unity since the ACF is normalized to unity. By
using a $\chi ^2$ minimization procedure, we obtained parameters
$C_1$, $C_2$ and $C_3$. The fitting procedure was a
search for the three non-linear parameters for the least-squared error
between model and data over the central region of ACF. The scintillation
parameters $\Delta \nu_{\rm d}$ and $\Delta t_{\rm d}$ are calculated
as:
\begin{align}
    \centering
    \Delta \nu _{\rm d} & = \sqrt {\ln 2/C_1} ,\\
    \Delta t_{\rm d} & = \sqrt {1/C_3}.
\end{align}

Following \citet{brg99a}, we use $dt/d\nu$ to describe the orientation
of the elliptical Gaussian. This is proportional to the refractive
scattering angle $\theta_{\rm r}$, and is given by:
\begin{equation}
    \centering
    \frac{dt}{d\nu} =  - \Big(\frac{C_2}{2C_3}\Big)
\end{equation}.

\begin{figure*}
    \centerline{\psfig{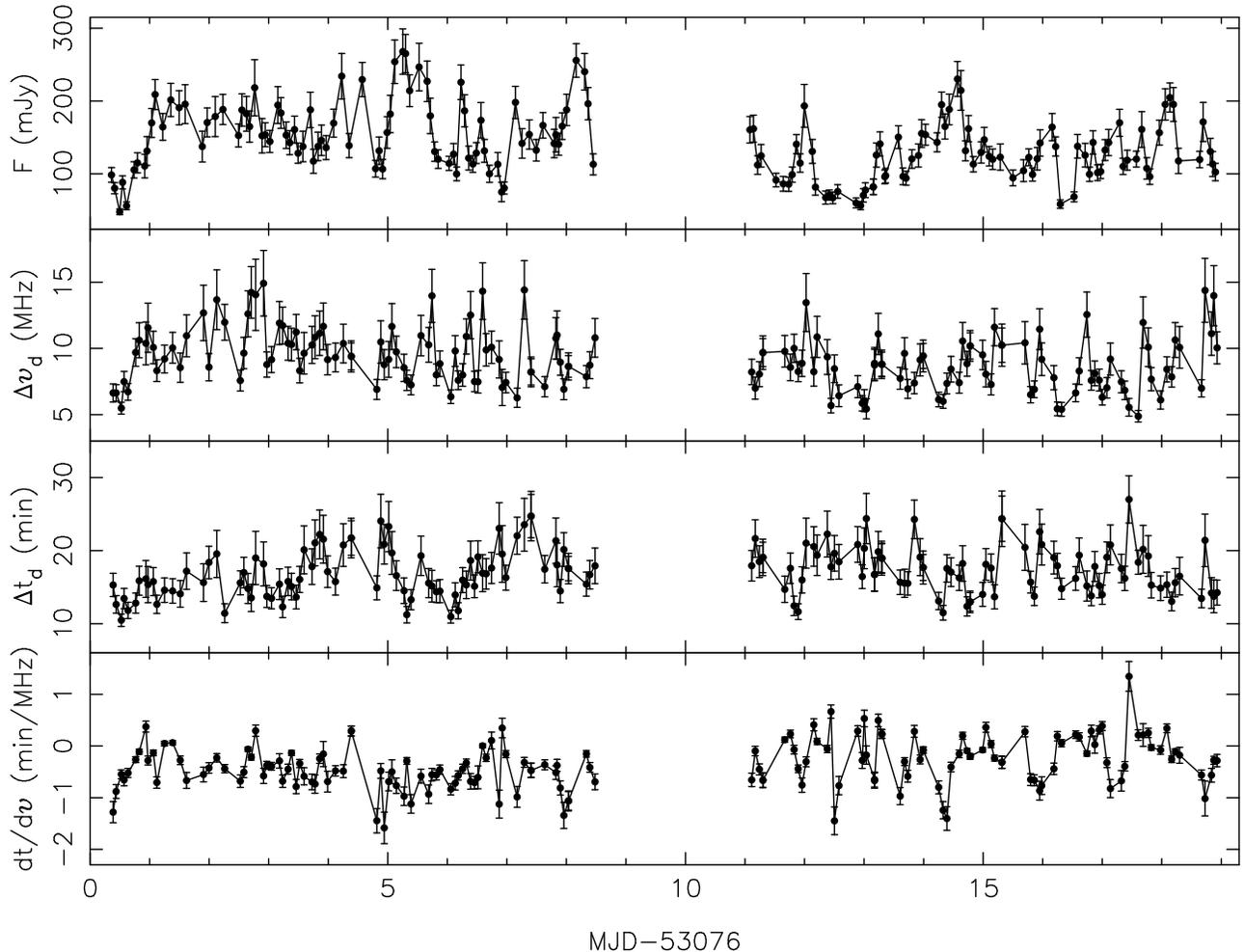}}
    \caption{Time series of flux density ($F$), decorrelation bandwidth
    ($\Delta \nu_{\rm d}$), diffractive time-scale ($\Delta t_{\rm
        d}$) and band drift rate ($d \rm t/d\nu$) . The gap in the plots represents
    the interruption of the observations for about three days. }
\label{fg:para}
\end{figure*}

From the $\chi^2$ analysis, we also obtained the uncertainties for
$C_1$, $C_2$ and $C_3$. Using these uncertainties
added in quadrature with the statistical error
(Equation~\ref{eq:staerr}), we calculate the errors of $\Delta
\nu_{\rm d}$, $\Delta t_{\rm d}$ and $dt/d\nu$. Since $dt/d\nu$ has
zero mean, we applied the statistical error to $(|C_2|+\sqrt{C_1
  C_3})/2$ rather than to $C_2$ itself. For flux density, measurement
errors are estimated from the baseline noise in the mean pulse
profiles. Time variations of $F$, $\Delta \nu_{\rm d}$, $\Delta t_{\rm
  d}$ and $dt/d\nu$ along with their uncertainties are shown in
Fig.~\ref{fg:para}. These parameters show significant variations with
time which can be explained as an RISS effect as discussed in
Section~\ref{sec:interpretation}.

\subsection{Derived Scattering Parameters}
Given a power-law spectrum of electron density fluctuations
(Equation~\ref{eq:spectrum}), the average scattering level along the
line-of-sight, $C_{\rm n}^2$, for a Kolmogorov spectrum is given by
\begin{equation}
    C_{\rm n}^2=2 \times 10^{ - 6} (\nu_{\rm MHz} )^{11/3} (D_{\rm pc} )^{ -11/6}\\
    (\Delta \nu _{\rm d,kHz} )^{ - 5/6}\;\; {\rm m}^{-20/3}
\end{equation}
\citep{cwb85}. 

Another quantitative measurement of the strength of scattering is the
parameter $u$ which is defined as the ratio of the Fresnel scale
$s_{\rm F}$ to the coherence scale $s_d =(\kappa \theta_{\rm
  d})^{-1}$ \citep{ric90}, where $\theta_{\rm d}$ is the diffractive
scattering angle.  For strong scattering, $u \gg 1$. In terms of the
diffractive scintillation parameter $\Delta \nu_{\rm d}$:
\begin{equation}
    u \approx \left(\frac{{2\nu }}{{\Delta \nu _{\rm d} }}\right)^{0.5}.
\end{equation}

The diffractive timescale depends on the velocity of the ray path
across the scattering screen. Assuming a thin screen and that
the pulsar velocity dominates, an estimate for the
pulsar transverse velocity based on the scintillation parameters is:
\begin{equation}
\label{eq:v}
    V_{\rm iss}  = A_{\rm V} \left(\frac{{\sqrt {x D_{\rm kpc} \;\Delta\nu_{\rm d,MHz} } }}
    {{\nu_{\rm GHz}\; \Delta t_{\rm d,s} }}\right)\;\;{\rm km s}^{-1},
\end{equation}
where $x = r_{\rm os}/r_{\rm ps}$ is the ratio of
the observer-screen distance to the pulsar-screen distance and the
proportionality constant $A_{\rm V}=3.85\times10^4$ \citep{grl94}.

In Table~\ref{tb:param} we give some basic parameters for the pulsar
and observations as well as mean values for the basic scattering
parameters. The pulsar distance is based on a parallax measurement by
\citet{bbgt02} who also measured the pulsar proper motion,  $19.5 \pm
0.4$ mas yr$^{-1}$ which, with the distance, gives the transverse velocity
listed in the table.  The typical observation time for each dynamic spectrum and
the total number of such observations are given.

\begin{table}
    \centering
    \caption{Observation Parameters and Results}\label{tb:param}
    \begin{tabular}{lr}
    \hline
    Galactic longitude (deg) & 145.00 \\
    Galactic latitude (deg)  & $-1.22$ \\
    DM (pc cm$^{-3}$)    &    26.83    \\
    Distance (kpc)           &    $1.06\pm 0.12$    \\
    $V_{\rm pm}$ (km~s$^{-1}$)  &  $90  \pm 2$\\
    Observation time (min)   & 90 \\ \\
    Number of observations       &    168      \\
    $\langle \Delta \nu_{\rm d}\rangle$ (MHz)                  &  $9.2\pm2.2$    \\
    $\langle \Delta t_{\rm d}\rangle$ (min)                 &  $17.1\pm3.3$   \\
    $\langle d \rm t/d\nu\rangle$ ($\rm min~MHz^{-1}$) &  $-0.36\pm0.46$   \\
    $\Delta t_{\rm r}$ (h)                                 &  $8\pm2$  \\
    $C_{\rm n}^2$ ($10^{-4}{\rm m}^{-20/3}$)       &  $13.8\pm3.9$   \\
    $u$                                                  &  $18.3\pm2.2$   \\
    $V_{\rm iss}$ (km~s$^{-1}$)                    &  $73\pm17$   \\
    \hline
    \end{tabular}
\end{table}

In the lower part of the table we give the mean values and rms scatter
for the decorrelation bandwidth ($\Delta \nu_{\rm d}$), scintillation
time-scale ($\Delta t_{\rm d}$) and the slope parameter ($dt/d\nu$)
derived from the ACF analysis. We note that the scintillation
timescale $\Delta t_{\rm d}$ is much less than the observation time
(90 min), so the effects of finite observation time discussed by
\citep{ssh+00} should be minimal. The predicted refractive timescale,
$\Delta t_{\rm r}$ based on the diffractive parameters
(Equation~\ref{eq:tref}), is about 4~d, but the observed value
(Section~\ref{sec:ref-time}) is much less than that. The derived scattering level,
$C_{\rm n}^2$, and scattering strength, $u$, both indicate that
scattering is strong along the path to PSR B0329+54. For the quoted
scintillation velocity, a centrally located screen ($x=1$) has been
assumed. Our value is somewhat lower than the proper motion
velocity. It is much lower than the value given for this pulsar by
\citet{grl94} based on 408 MHz observations, but they assumed a pulsar
distance of 2.3 kpc. For the parallax distance of $\sim 1.0$ kpc,
their velocity becomes $105\pm17$ km~s$^{-1}$, consistent with the
proper motion velocity. From their 327 MHz observations \citet{brg99a}
derive a scintillation velocity of $186\pm 17$ km~s$^{-1}$ based on an
assumed pulsar distance of 1.43 kpc. Scaled with the parallax
distance, this becomes $155\pm 14$ km~s$^{-1}$. While these results
are not too discordant, it is clear that there are variations on
timescales much longer than the nominal refractive time and also that
the frequency scaling of the diffractive parameters does not precisely
follow the Kolmogorov prediction \citep[cf.][]{wmj+05}.

\section{Interpretation of Results}\label{sec:interpretation}

\subsection{Cross-Correlations Between Scintillation Parameters}
\citet{rnb86} suggested that correlations should exist between
variations of the flux density $F$, decorrelation bandwidth
$\Delta \nu_{\rm d}$, and diffractive timescale $\Delta t_{\rm
d}$. These predictions assume a single-phase screen and a simple
power-law description for the electron density fluctuations.
In Table~\ref{tb:cross} we list the predicted correlation coefficients
for turbulence spectra of $\beta=$11/3, 4 and 4.3 \citep{rnb86}.
\begin{table*}
   \caption{The theoretical predication and observational results of
     cross-correlation coefficients.}
    \label{tb:cross}
    \begin{minipage}{150mm}
    \begin{tabular}{lcccc}
    \hline
  & &\{$\Delta \nu_{\rm d}$, $\Delta t_{\rm d}$\} &\{$\Delta \nu_{\rm d}$, $F$\}  &\{$\Delta t_{\rm d}$, $F$\}\\
    \hline
     RNB86    &  $\beta=11/3$    &$0.75$  &$-0.76$  &$-0.50$\\
                                              &  $\beta=4$       &$0.77$  &$-0.80$  &$-0.58$\\
                                              &  $\beta=4.3$     &$0.79$  &$-0.84$  &$-0.64$\\
    \hline
     SFM96  &  $r$   &$0.65$  &$-0.40$  &$-0.62$\\
               &  Spearman $r_{\rm s}$  &$0.54$  &$-0.41$ &$-0.68$ \\
    \hline
    BRG99b       &  Spearman $r_{\rm s}$  &$0.04$  &$0.45$  &$-0.61$ \\
    \hline
    This work    &  Spearman $r_{\rm s}$  &$0.14$  &$0.30$  &$-0.07$ \\
                 &  95\% confidence interval &   $-0.03 \sim 0.30$ &  $0.17 \sim 0.44$   & $-0.22 \sim 0.08$ \\
    \hline
    \end{tabular}
    \end{minipage}
\end{table*}

\citet{sfm96} and \citet{brg99b} studied the correlations between the
variations of $F$, $\Delta \nu_{\rm d}$ and $\Delta t_{\rm d}$ for PSR
B0329+54. Both papers use the Spearman rank-order correlation coefficient
($r_{\rm s}$) to describe the correlation between the parameters.
$r_{\rm s}$ is given by:
\begin{equation}
    r_{\rm s}  = \frac{{\sum\limits_{\rm i} {(R_{\rm i}  - \overline R )(S_{\rm i}  - \overline S )} }}
    {{\sqrt {\sum\limits_{\rm i} {(R_{\rm i}  - \overline R )} ^2 } 
              \sqrt {\sum\limits_{\rm i} {(S_{\rm i}  - \overline S )} ^2 } }}
\label{eq:rank}
\end{equation}
where $R_{\rm i}$ and $S_{\rm i}$ are the ranks of the two quantities
$x_{\rm i}$ and $y_{\rm i}$ for which the correlation coefficient is
computed and the summation is carried out over the total number of the
data points.  $\overline R$ and $\overline S$ represent the average
values of $R_{\rm i}$ and $S_{\rm i}$. This method is less sensitive
to the outlying points than the usual correlation coefficient $r$,
which is computed with the same equation but with the actual data
values rather than their rank. $r_{\rm s}$ and its confidence interval
are derived using the `bootstrap' method as follows: assume we have an
original object O with a set of N elements. We construct a new list
with the same number of N elements from the original list by randomly
picking elements from the list. Any one element from the list can be
picked any number of times. The process is repeated 10000 times to
give new series \{$O_1$, $O_2$....$O_{\rm 10000}$\} from
which we obtain a statistical distribution of $r_s$. The mean value
and 95\% confidence limits are derived from this distribution. 
\begin{figure*}
    \centerline{\psfig{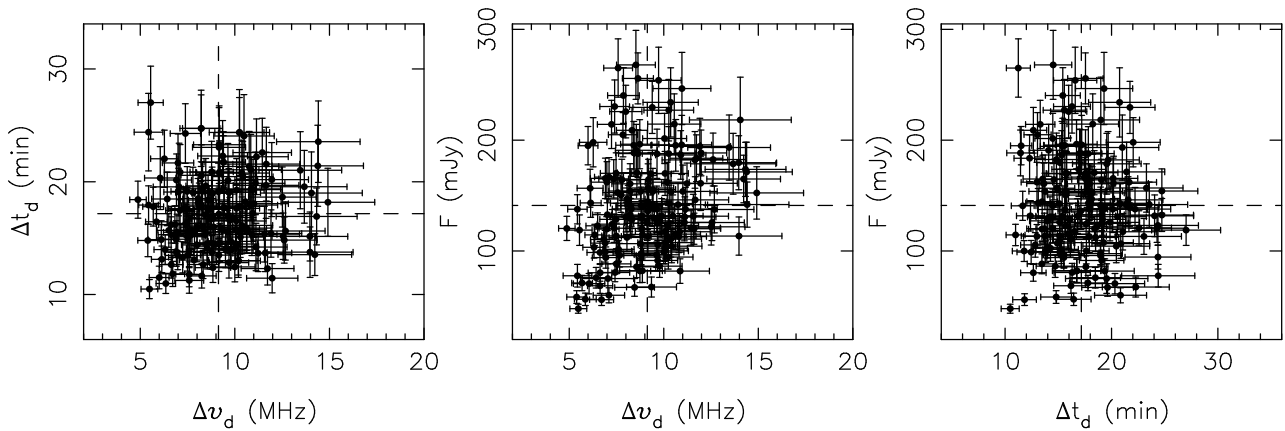}}
    \caption{Scatter plots of $\Delta \nu_{\rm d}$ vs $\Delta t_{\rm d}$,
    $\Delta \nu_{\rm d}$ vs $F$ and $\Delta t_{\rm d}$ vs $F$, respectively.}
\label{fg:correlation}
\end{figure*}
The results for $r_{\rm s}$ are presented in Table~\ref{tb:cross},
along with results from \citet{sfm96} and \citet{brg99b} for
comparison. 

Although \citet{sfm96} found significant cross-correlations between
the scintillation parameters which were approximately in accord with
the theoretical predictions, neither \citet{brg99b} nor our
observations show such correlations. As Fig.~\ref{fg:correlation}
shows, little or no correlation is observed between $\Delta \nu_{\rm
d}$ and $\Delta t_{\rm d}$ and the correlation between $\Delta
\nu_{\rm d}$ and $F$ even has the opposite sign to the theoretical
prediction. In contrast to the other studies, we observe only a very
weak correlation between $\Delta t_{\rm d}$ and $F$. The observed
correlations are clearly not well described by the thin-screen model
of \citet{rnb86}. Their predictions are not strongly dependent on the
assumed slope of the fluctuation spectrum so other factors, for
example, an extended scattering region, must be important. However,
there have been no predictions for the expected correlations in this
case.

\subsection{Structure function analysis}
We use a structure function analysis to study the refractive
variations in the flux density and the diffractive timescale and
decorrelation bandwidth. The structure function is defined to be:
\begin{equation}
\label{eq:struct}
D(n) = \frac{1}{{ \langle F \rangle ^2 N(n)}}\sum\limits_{i = 1}
{w(i)w(i+n)[F(i) - F(i + n)]^2 }
\end{equation}
where $D(n)$ is the structure function at lag of n units of the
observation time (90 min), $F(i)$ is the flux density value for the
$i$th observation, $w(i) = 1$ if $F(i)$ exists and zero otherwise,
$\langle F\rangle$ is the mean flux density, and $N(n)$ is the number
of products in $D(n)$ for which $w(i)w(i+n) \neq 0$. Typically, the
structure function has three regimes: a noise regime at small lags, a
structure regime characterized by a linear slope on a log-log plot and
a saturation regime where it flattens out at large lags. The shape of
the structure function may be used to derive the refractive modulation
index $m_{\rm r}$ and the refractive scintillation time-scale $\Delta
t_{\rm r}$. The modulation index $m_{\rm r}$ is expressed as $m_{\rm
  r}=\sqrt {D(\infty )/2}$, where $D(\infty)$ is the saturation value
of the structure function, and the refractive time-scale $\Delta t_{\rm
  r}$ is the time lag at which the structure function reaches half of
its saturation value. The logarithmic slope of the structure function,
$\gamma$, is defined in the linear structure region. Structure
functions for the refractive variations in diffractive timescale and
bandwidth may be similarly defined.

\begin{figure}
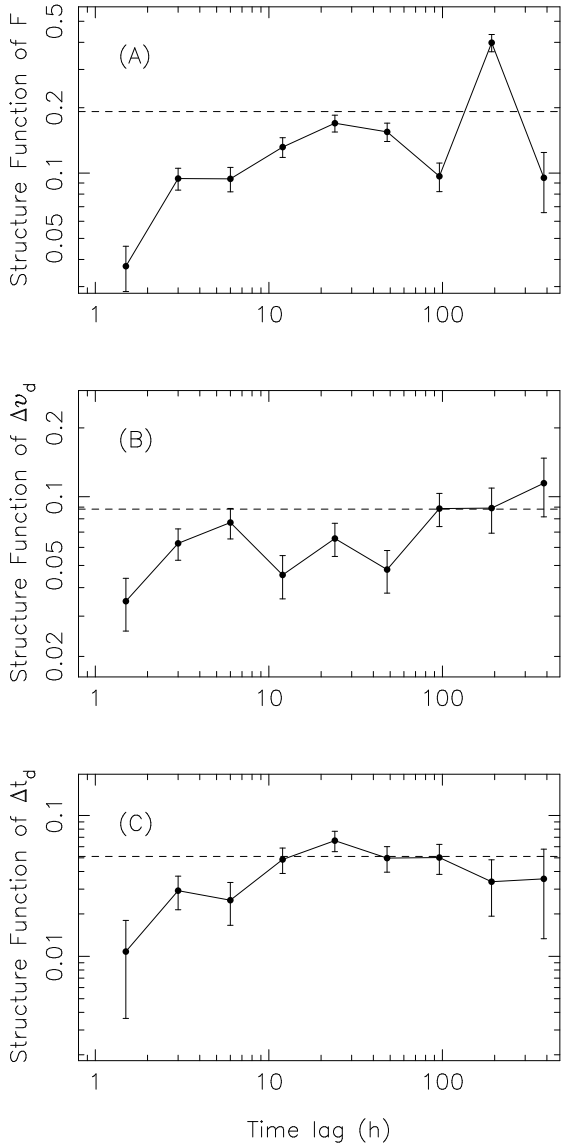

\begin{center}
\begin{tabular}{c}
    \mbox{\psfig{file=struc_flux.ps,height=50mm,angle=270}}\\
    \mbox{\psfig{file=struc_freq.ps,height=50mm,angle=270}}\\
    \mbox{\psfig{file=struc_time.ps,height=50mm,angle=270}}
\end{tabular}
\caption{Log-log plots of structure functions for observed variations
  of: (A) Flux density $F$, (B) decorrelation band-width $\Delta
  \nu_{\rm d}$ and (C) diffractive time-scale $\Delta t_{\rm d}$. The
  plotted structure functions have been corrected for the estimated
  noise contribution. The horizontal dashed lines are at the level
  corresponding to the observed noise-corrected modulation index. }
\label{fg:struct}
\end{center}
\end{figure}

Structure functions must be corrected for the effects of noise in the
time series. The noise contribution to the structure function values,
$D_{\rm noise} = 2\sigma_{\rm noise}^2 = 2(\sigma_{\rm est}^2 + \sigma_{\rm meas}^2)$,
where $\sigma_{\rm est}$ is the statistical estimation error resulting
from the finite number of scintles in the dynamic spectrum
(Equation~\ref{eq:staerr}), $\sigma_{\rm meas}$ is the mean fractional
uncertainty in the measured parameter. In Table~\ref{tb:noise} we list
measured values of $\sigma_{\rm meas}$, $\sigma_{\rm est}$, $\sigma_{\rm noise}$
and $D_{\rm noise}$ for the three quantities, along with the uncorrected
value of $D(1)$, the 90-min lag value. Corrected values of the derived
structure functions, that is, with $D_{\rm noise}$ subtracted, are shown
in Fig.~\ref{fg:struct}. Uncertainties on the structure function
values are given by $\sigma_D(n) = \sigma_{\rm noise}[8 D(n) /
  N(n)]^{1/2}$ \citep{sc90a}. In all cases, the corrected value of
$D(1)$ is below the extrapolation of the linear part of the structure
function, indicating that we may have somewhat over-estimated the
noise contribution.

\begin{table}
\caption{Structure function noise estimates}\label{tb:noise}
\begin{tabular}{lccccc}
\hline
 Parameter & $\sigma_{\rm meas}$ & $\sigma_{\rm est}$ & $\sigma_{\rm
   noise}$ & $D_{\rm noise}$ & $D(1)$ \\   \hline
$F$   &  0.045  &  0.110  & 0.119 & 0.028 & 0.065   \\
$\Delta\nu_{\rm d}$ &  0.058  & 0.110 & 0.124 & 0.030 & 0.065  \\ 
$\Delta t_{\rm d}$  &  0.058  & 0.110 & 0.124 & 0.030 & 0.041  \\  \hline
\end{tabular}
\end{table}

\subsection{Modulation Indices of Scintillation Parameters}
In strong scattering, the diffractive parameters are modulated by
refractive effects.  \citet{rnb86} gave relations for the expected
modulation indices for refractive variations of flux density ($m_{\rm
  r}$), diffractive bandwidth ($m_{\rm b}$) and diffractive timescale
($m_{\rm t}$) for power-law fluctuation spectra with $\beta=11/3$, 4
and 4.3. For the steeper spectra, the modulation indices are largely
independent of scattering strength and frequency. They found that the
depth of modulation is lowest for the Kolmogorov spectrum and larger
for larger values of $\beta$. Predicted modulation indices for PSR
B0329+54 are given in columns 2 -- 4 of Table~\ref{tb:mod} . Based on
the \citet{rnb86} theory, \citet{bgr99} gave the modulation indices of
$F$, $\Delta \nu_{\rm d}$, and $\Delta t_{\rm d}$ for $\beta=4$ and
$\beta=4.3$. We list them in column 3 \& 4 of
Table~\ref{tb:mod}. 

The noise-corrected  modulation indices may be computed directly from
the time series using 
\begin{equation}
m = \sqrt{\sigma^2 - \sigma_{\rm noise}^2}
\end{equation}
where $\sigma$ is the observed rms fluctuation relative to the
mean. The modulation index has an uncertainty given by
\begin{equation}
\sigma_m = \frac{1}{2m} \left(\frac{\sigma^4}{N_{\rm scint}} + 
          \frac{\sigma_{\rm noise}^4}{N_{\rm obs}}\right)^{1/2}
\end{equation}
where $N_{\rm scint}$ is approximately the data span divided by the
refractive timescale and $N_{\rm obs}$ is the number of independent
observations.  Using the noise estimates given in
Table~\ref{tb:noise}, the derived modulation indices and their
uncertainties are given in column 5 of Table~\ref{tb:mod}. The derived
values are consistently larger than the predicted values for a
Kolmogorov spectrum \citep[cf.][]{ssh+00}, indicating a $\beta \sim
4.0$.

\begin{table}
    \caption{Modulation indices for DISS parameters}\label{tb:mod}
    \begin{tabular}{lcccc}
    \hline
        \ & &Prediction & &Our\\
        \cline{2-4}
        \ &$\beta=11/3$ &$\beta=4$&$\beta=4.3$&Observation \\
        \hline
        $m_{\rm r}$    &0.22   &0.38      &0.55    & $0.31\pm 0.03$ \\
        $m_{\rm b}$    &0.15   &0.35      &0.57    & $0.21\pm 0.02$ \\
        $m_{\rm t}$    &0.07   &0.17      &0.25    & $0.16\pm 0.02$ \\
    \hline
    \end{tabular}
\end{table}

Another possible intepretation of the relatively large observed
modulation indices is that the scattering medium is extended so that
the thin-screen approximation is not valid. The flux density
modulation index for a statistically uniform scattering medium with a
Kolmogorov spectrum ($\beta = 11/3$) covering the whole path to the
pulsar has been discussed by \citet{cfrc87}, \citet{grc93},
\citet{ssh+00} and \citet{sss98} with the result
\begin{equation}
m_{\rm r} \approx 1.05 (\Delta\nu_d / \nu)^{0.167}.
\end{equation}
For our observations of PSR B0329+54, this relation gives a predicted
$m_{\rm r} \approx 0.45$, larger than the observed value. This implies
that, as an alternative to a steeper spectrum, the observed modulation
index could be accounted for by an extended Kolmogorov scattering
medium covering just part of the path.

\subsection{Refractive Timescales} \label{sec:ref-time}
The horizontal dashed lines on the structure function plots
(Fig.~\ref{fg:struct}) are at $D(\infty) = 2 m^2$, where $m$ is the
observed modulation index given in in Table~\ref{tb:mod}. The
refractive timescale $\Delta t_{\rm r}$ is defined to be the lag at
which the structure function falls to $0.5 D(\infty)$. In the flux
density case, this is at $8\pm2$ h. The structure functions for the
diffractive bandwidth and timescale are less well defined, but clearly
also imply a short refractive timescale. The measured timescale is
much less than the predicted value of about 96~h based on a central
thin-screen model and the observed diffractive timescale
(Table~\ref{tb:param}). For an extended scattering medium with a
Kolmogorov spectrum, evaluating the equations of \citet{sss98} with
practical units gives
\begin{equation}
\Delta t_{\rm r,h} = 278 \frac{R_{\rm pc}^{1/2}}{V_{\rm km\,s^{-1}} \;
  \Delta\nu_{\rm d,MHz}^{1/2}}
\end{equation} 
Substituting the parameters from our observations of PSR B0329+54
gives a value of about 32 hours, again much larger than the observed
value. 

\citet{sss98} also consider the case of a thin screen at an arbitrary
location along the path and show that $\Delta t_{\rm r}$ is
proportional to the pulsar-screen distance ($r_{\rm ps}$) whereas
$\Delta t_{\rm d}$ is proportional to $R/(R - r_{\rm ps})$. Therefore,
if $r_{\rm ps} << R$, refractive times are much reduced, whereas
$\Delta t_{\rm d}$ is only weakly dependent on $r_{\rm ps}$. This
breaks the approximate proportionality of $\Delta t_{\rm r}$ and
$\Delta t_{\rm d}$ (Equation~\ref{eq:tref}) which applies to both a
central thin screen and an extended medium. The scintillation arc
observations of \citet{ps06a} suggest that most of the scattering for
PSR B0329+54 occurs relatively close to the pulsar, supporting this
interpretation of the short refractive timescale.

\subsection{Structure Function Slope}
The slope of the linear part of the structure function, between the
noise and saturation regimes, is related to the power-law index of the
fluctuation spectrum. For a single thin screen, the slope should be
2.0 for $\beta \le 4$ \citep{rnb86,sss98}. For an extended screen, the
predicted slope is $\beta -3$ \citep{sss98}. Although the observed
slopes are not very well determined, they are clearly $\ll 2.0$. The
measured value for the flux density structure function slope is
$0.4\pm 0.2$ and the values for the other two structure functions are
similar. This is clearly inconsistent with scattering by a single thin
screen. As we have argued above, there are good arguments supporting
the idea that the scattering screen in this direction is extended; in
this case, the implied value of $\beta = 3.4 \pm 0.2$, somewhat less
than but marginally consistent with the Kolmogorov value. Based on the
thin-screen model, observed diffractive bandwidths and a refractive
interpretation of band slopes ($dt/d\nu$), \citet{brg99a} and
\citet{wmj+05} also obtain $\beta$ values less than the Kolmogorov
value for PSR B0329+54 and several other pulsars. Our result also
agrees well with that from \citet{ssh+00} who obtain a structure
function slope of $0.5\pm 0.1$ from their flux density monitoring of
this pulsar.  However, it contrasts with that of \citet{sss+03} where
a structure function slope $\sim 1.5$ is derived from observations at
several different frequencies and interpreted in terms of weak plasma
turbulence, again with a Kolmogorov spectrum.

\section{Conclusions}
Quasi-continuous observations of PSR B0329+54 at 1540 MHz over 20 days
have been analysed to investigate the refractive modulations of the
pulsar flux density and diffractive scintillation parameters. The data
set was split into more than 150 individual observations, each of
90-min duration, and two-dimensional auto-correlations and secondary
spectra computed. Time series of the pulsar flux density ($F$),
decorrelation bandwidth ($\Delta \nu_{\rm d}$), diffractive
scintillation time-scale ($\Delta t_{\rm d}$) and drift rate of
features in the dynamic spectra ($dt/d\nu$) clearly show refractive
modulation, although no evidence for arc structure is seen in the
secondary spectra.

Observed cross-correlations between variations in the flux density and
diffractive parameters were much smaller than predictions based on the
thin-screen model. In one case the sign of the correlation was
opposite to the predicted value. Similar results have been obtained by
other observers although different observations appear to give
conflicting results. 

In accordance with previous work, observed modulation indices are
greater than predicted for a thin screen with a Kolmogorov fluctuation
spectrum.  This could be accounted for either by a steeper fluctuation
spectrum ($\beta \sim 4$) or by scattering in an extended medium. The
predicted modulation index for a scattering medium covering the whole
path to the pulsar is greater than that observed, suggesting that the
scattering medium, while extended, does not cover the whole
path. Structure functions derived from the observations indicate a
short refractive timescale, $8\pm 2$~h, much less than predicted from
the thin screen model, and have a relatively flat slope, $0.4\pm 0.2$,
again inconsistent with scattering by a thin screen.

The observed modulation indices, structure function slopes and short
refractive timescales all are consistent with scattering by an
extended region with a Kolmogorov fluctuation spectrum which is
concentrated toward the pulsar. This idea is supported by recent
high-sensitivity observations of PSR B0329+54 by \citet{ps06a} which
show indistinct scintillation arcs corresponding to extended scattering regions
relatively close to the pulsar.

\section*{Acknowledgments}
We would like to thank B.~J.~Rickett and W.~A.~Coles for helpful discussions.
This work is supported by the Key Directional Project of CAS and NNSFC
under the project 10173020 and 10673021.


\end{document}